\documentclass{iaus238}
\usepackage{graphicx}
\newcommand{\msun}{M_{\odot}}

\title[Soft X-ray spectrum of PG1211+143] 
{The soft X-ray spectrum of PG1211+143}

\author[K.~A. Pounds]   
{K.~A. Pounds,$^1$ K.~L. Page$^1$ \and J.~N. Reeves$^2$}

\affiliation{$^1$~Department of Physics and Astronomy, University of Leicester, UK\\[\affilskip]
$^2$~NASA Goddard Space Flight Center, Code 663, Greenbelt, MD 20771, USA \\ }

\pubyear{2006}
\volume{238}  
\pagerange{001--999}
\date{??? and in revised form ???}
\jname{Black Holes: from Stars to Galaxies -- across the Range of Masses}
\editors{V. Karas \& G. Matt, eds.}
\begin{document}

\maketitle

\begin{abstract}
The narrow line QSO PG1211+143 has been a focus of recent attempts to
understand the soft excess in AGN, while the 2001 {\it XMM-Newton}
observation of this luminous AGN also provided evidence for a massive
and energetic outflow. Here we consider a physical link between the
energetic outflow and the variable soft excess.
\end{abstract}

\firstsection % if your document starts with a section,
              % remove some space above using this command.
\section{Introduction}

One of the most important new results emerging from {\it XMM-Newton} and
{\it Chandra} observations of AGN is the evidence for sub-relativistic
outflows from a number of  luminous quasars.  For two of these,
APM08279+5255 (Chartas \etal\ 2002) and PG1115+080 (Chartas \etal\
2003), both BAL quasars, the immediate inference was to associate the
X-ray columns with the high velocity gas seen in optical and UV spectra
of these unusual objects. In such cases the high columns might result
from viewing the continuum source through a wind being blown
tangentially off the accretion disc (Proga \etal\ 2000).

However, in 2 further cases, PG1211+143 (Pounds \etal\ 2003) and PDS456
(Reeves \etal\ 2003), a moderate and a high luminosity QSO,
respectively, evidence of a high velocity -- and hence energetic --
outflow suggested this could be a common property of luminous, or
perhaps more specifically, of high accretion rate quasars (King and
Pounds 2003). Unless highly collimated, such outflows will form a
substantial  component of the mass and energy budgets in luminous AGN,
while having important implications for metal enrichment of the
intergalactic medium and for the feedback mechanism implied by the
correlation of black hole and galactic bulge masses (King 2005).

The {\it XMM-Newton} observation of PG1211+143 in 2001 has also been
used recently to explore the physical origin of the `soft excess' in the
X-ray spectra of many AGN, following doubts raised about Comptonisation
models (Gierlinski and Done 2004). Alternative descriptions have been
proposed, invoking strong reflection  from a highly ionised accretion
disc (Crummy \etal\ 2006) and absorption in high velocity ionised gas
(Schurch and Done 2006, Chevallier \etal\ 2006). Here we also examine
EPIC data from a second observation of PG1211+143 in 2004, where a
conventional plot shows a weaker but `hotter' soft excess (Figure 1L).

A direct comparison of the EPIC data from 2001 and 2004 (Figure 1R)
suggests the spectral difference is primarily due to a variable
`deficit' of flux at $\sim$0.7--2 keV, supporting the contention that the
spectral curvature near $\sim$1 keV is indeed an artefact of absorption
on a  steeper underlying continuum. However, if the flux deficit is due
to photoionised absorption,  it is interesting that the hard X-ray flux
-- which should be a good proxy for the ionising luminosity -- is
essentially unchanged, suggesting that a reduced covering factor (or
column density) is the cause of the weaker `soft excess' in 2004.  Here
we suggest an alternative possibility (explored more fully in Pounds and
Reeves 2006), that a variable continuum component partly `fills in' the
absorption flux deficit in the 2004 spectrum. While contributing
directly to the strong soft X-ray flux in PG1211+143, the  additional
soft continuum component also explains the lack of strong `narrow'
absorption lines in the soft X-ray band (Kaspi and Behar 2006), in
conflict with the relatively strong, blue-shifted absorption lines  in
the EPIC data (Pounds and Page 2006). 

We speculate that the additional, steep power law continuum is powered by
the energetic high velocity outflow. 

\begin{figure}
\rotatebox{-90}
{\includegraphics[width=.18\textheight]{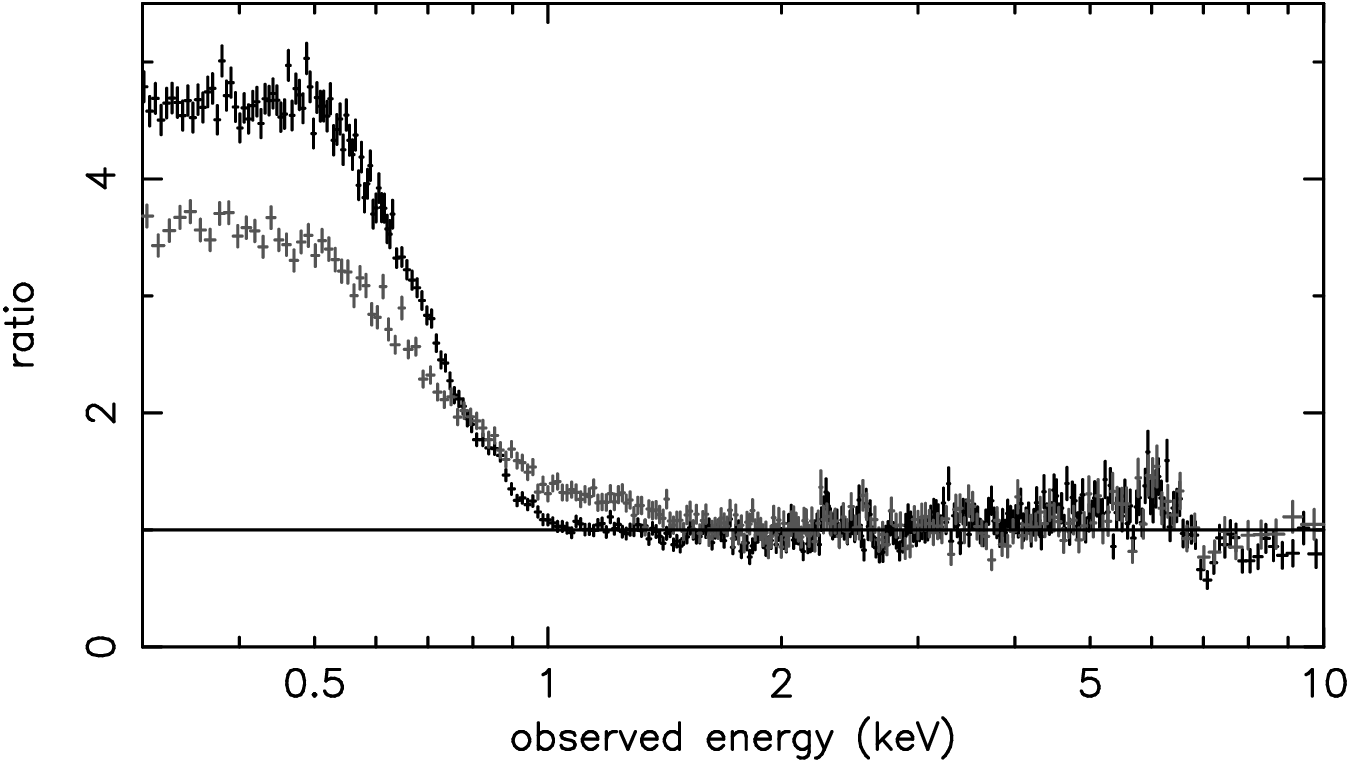}}
\rotatebox{-90}
{\includegraphics[width=.18\textheight]{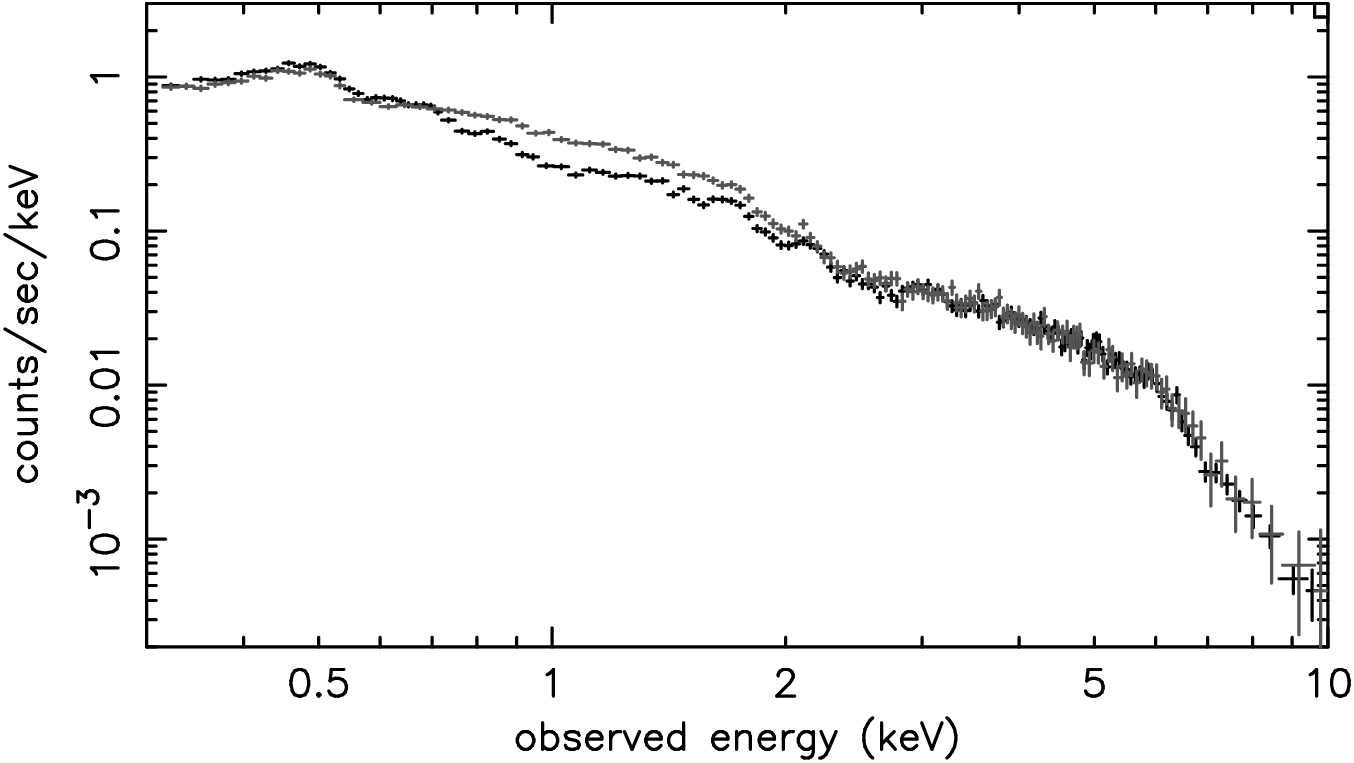}} 
\caption{Left panel: conventional plot of the 2001 (black) and 2004
(grey) EPIC pn spectra showing a strong soft excess sitting above the
hard power law. On this view it appears that the soft excess in 2004 is
weaker but `hotter'.  Right panel: direct comparison of the
background-subtracted EPIC data  showing the spectral difference  to be
due to an increase in flux at $\sim$0.7--2 keV and a (less obvious, but
significant) decrease at $\sim$0.4--0.7 keV in 2004 (grey).}
\label{l}
\end{figure}

\section{Confirming an energetic outflow in PG1211+143}
PG1211+143 (at z=0.0809) is a prototypical `Big Blue Bump' quasar, and
one of the brightest AGN in the soft X-ray band (Elvis \etal\ 1991). It
is also classified (Kaspi \etal\ 2000) as a Narrow Line Seyfert 1 galaxy
(FWHM H$\beta$ 1800 km~s$^{-1}$), with a small central black hole mass 
($\sim4\times10^{7}$M$_{\odot}$) for an object of bolometric luminosity
$4\times10^{45}$erg~s$^{-1}$, implying that the accretion rate is close
to  the Eddington  limit.  

The initial $\sim$50 ks observation of PG1211+143 with XMM-Newton in
2001 revealed several blue-shifted absorption lines, interpreted by
Pounds \etal\ (2003) as evidence for a highly ionised outflow with a
velocity v$\sim$0.09c, a conclusion primarily based on identifying the
strongest $\sim$7 keV feature with FeXXVI Ly$\alpha$. As Kaspi and Behar
(2006) were unable to confirm this result in a careful ion-by-ion
modelling of the RGS data we have re-examined the higher signal-to-noise
EPIC data. That re-analysis resolved additional absorption lines in the
intermediate ($\sim$1--4~keV) energy band (Pounds and Page 2006),
strengthening the evidence for a high velocity outflow.  In total, 7
statistically significant absorption lines are identified, with the
strong $\sim$7 keV line re-asigned to He$\alpha$ of FeXXV, yielding an
{\it increased} outflow velocity  v$\sim$0.14$\pm$0.01c.

Assuming a spherically symmetric flow, the mass loss rate is of then of
order  $\dot M \sim 35b\msun$~yr$^{-1}$, where $b\leq$1 allows for the
collimation of the flow. Modelling of the broad-band spectrum of
PG1211+143 (Pounds and Reeves 2006) has quantified both the absorbed and
re-emitted fluxes for the ionised outflow, yielding a covering factor
CF$\sim$0.1 for the high velocity,  highly ionised outflow, and $\dot M$
$ \sim 3.5\msun$~yr$^{-1}$. This outflow rate compares with $\dot M_{\rm
Edd}$ = 1.6$\msun$~yr$^{-1}$ for a non-rotating SMBH of mass
$\sim$$4\times 10{^7}$$\msun$ (Kaspi \etal\ 2000) accreting at an
efficiency of 10\%.  The mechanical energy of the highly ionised outflow
is then $\sim10^{45}$ erg s$^{-1}$, easily sufficient to power a
substantial  component of the X-ray emission of PG1211+143.

\section{A second continuum component}
For most AGN the energetically dominant emission component can be
modelled by a power law of photon index $\Gamma\sim$1.9 (Nandra and
Pounds 1994),  visible in Type 1 objects over the $\sim$2--100 keV band.
This `primary' continuum component is widely believed to arise by
Comptonisation of UV and optical accretion disc photons in a high
temperature electron corona (Haardt and Marashi 1991). However, recent
studies of spectral variability in MCG--6-30-15 (Vaughan and Fabian 2004)
and 1H0419-57 (Pounds \etal\ 2004) have provided clear evidence of a
variable power law component of slope significantly  steeper than the
`canonical value' of $\Gamma\sim$1.9. 

We now suggest this softer continuum component is physically distinct
from the `primary' continuum in PG1211+143, having established that the
mechanical energy in the fast, highly ionised outflow  is ample to power
that additional continuum component, perhaps via internal shocks 
(analogous to the process suggested in Gamma Ray Bursts), or shock
heating of slower moving clouds providing the bulk of the continuum
opacity.   The most likely X-ray emission mechanism would again be
Comptonisation of optical-UV disc photons, with the steep power law
resulting from a lower equilibrium temperature in the dispersed `corona'
compared with that responsible for the primary power law continuum.

A separate continuum component is attractive in the present context
since, if absorption is to create the impression of a strong `soft
excess', then the underlying continuum must be steeper than observed
above $\sim$2 keV. 

\begin{figure}
\rotatebox{-90}
{\includegraphics[width=.18\textheight]{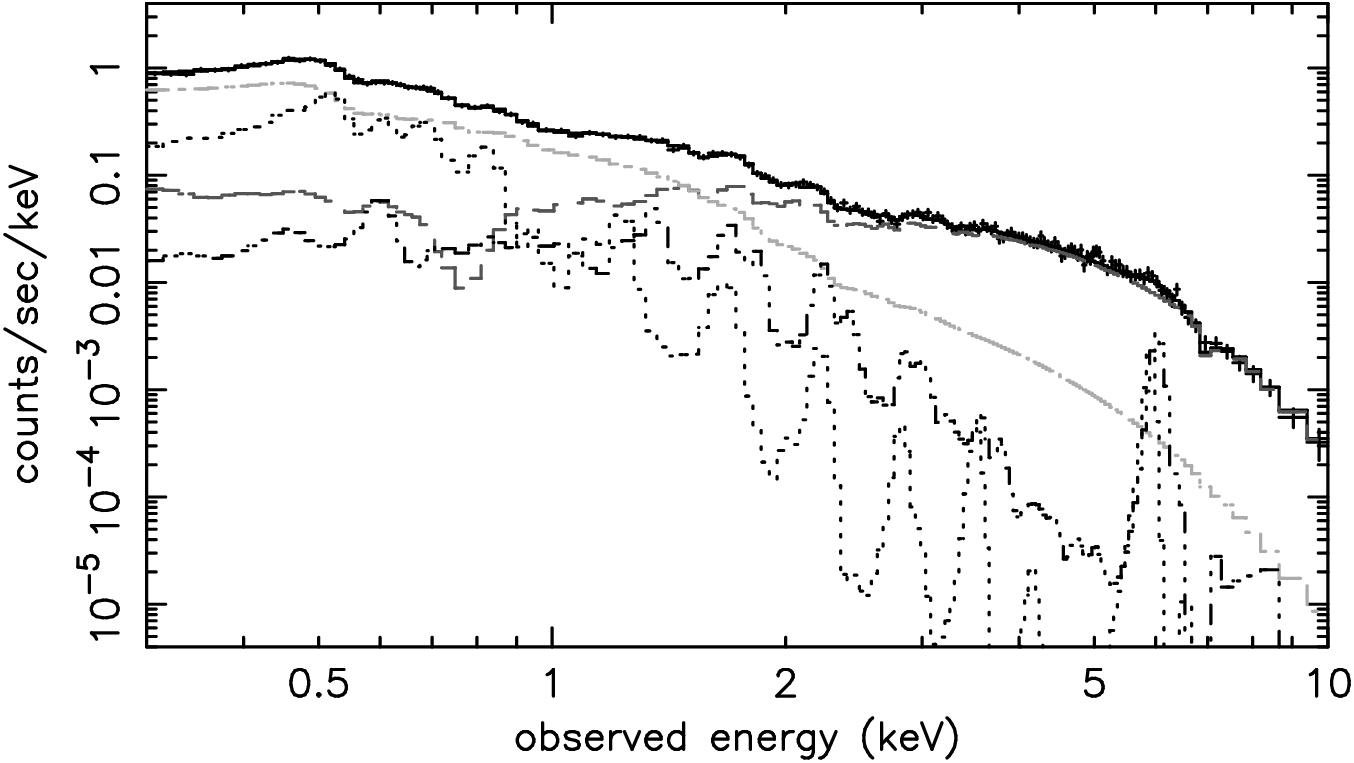}}
\rotatebox{-90}
{\includegraphics[width=.18\textheight]{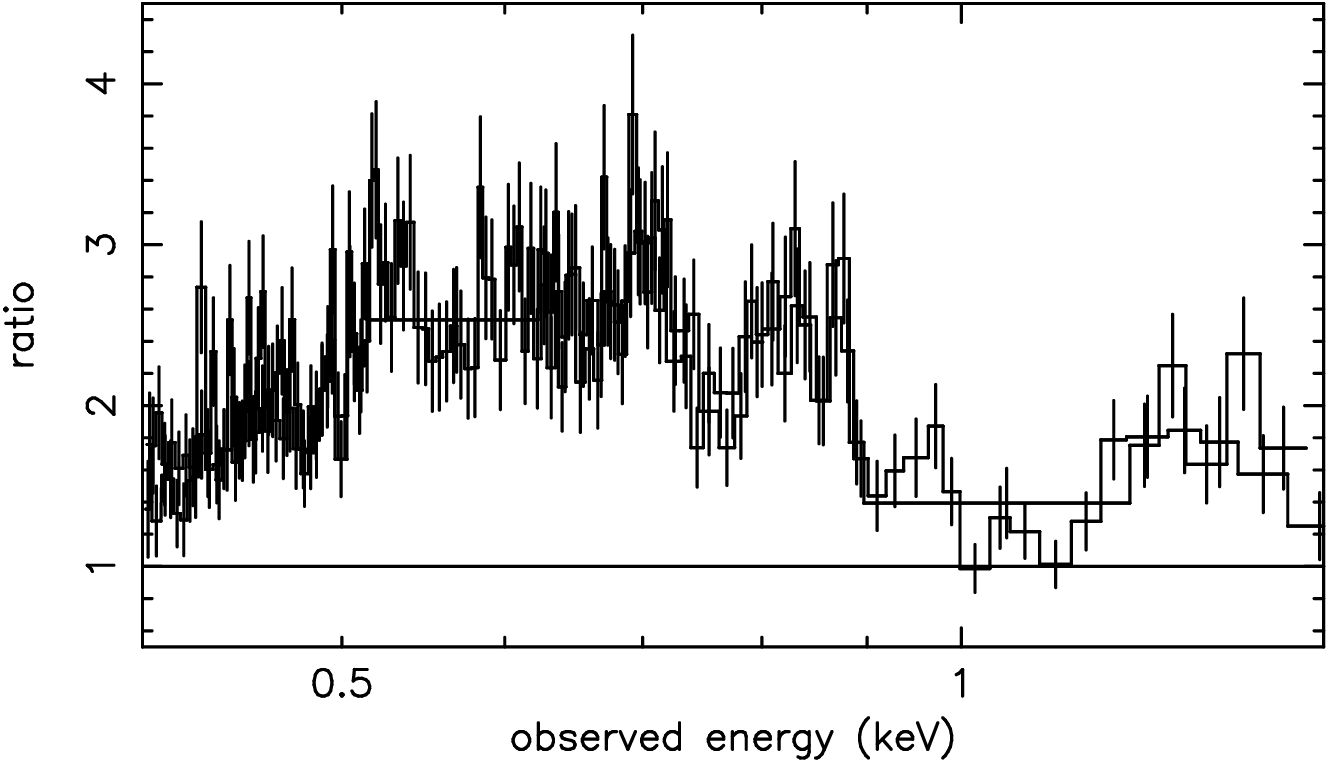}}
\caption{Left panel: deconvolution of the EPIC MOS spectrum of
PG1211+143 with the primary (dark grey) and secondary (light grey)
continua plus line emission from the  highly and moderately ionised
outflow (Pounds and Reeves 2006).  Right panel: structure in the soft
excess in the 2001 RGS data. XSTAR model fitting  shows this structure
is dominated by velocity-broadened resonance emission  lines of N,O,Ne
and Mg, together with a group of Fe-L lines at $\sim$0.7 keV.}
\label{3}
\end{figure}

\section{A new spectral model}
The evidence of absorption in highly ionised (narrow absorption lines)
and moderately ionised gas (low energy spectral curvature) suggests  a
new spectral model including 2 power law continua, both modified by
absorption, and re-emission from a 2-component photoionised gas.
Applying this model to the 2001 data from PG1211+143 produced a
statistically excellent fit (Pounds and Reeves 2006). 

Figure 2L illustrates the model components, showing that above $\sim$2
keV the spectrum is dominated by the `primary' power law, with a photon
index $\Gamma\sim$2.2, at the upper end of -- but consistent with -- the
`accepted' range for type 1 AGN spectra. At lower energies the 
secondary power law component ($\Gamma\sim$3.1) dominates, while below
$\sim$1 keV re-emission, particularly from the moderately ionised
absorber, becomes significant. 

Strong absorption of the primary power law is seen to be responsible for
the mid-band ($\sim$2--6 keV) spectral curvature, with ionised gas column
densities of N$_{H}\sim3-5\times 10^{22}$ cm$^{-2}$. In both pn and
MOS spectral fits the model shows a broad trough near $\sim$0.75--0.8
keV, due to Fe-L absorption. Of particular relevance in the present
context,  the luminosity of the steep, less absorbed, power law
component was $\sim6\times 10^{43}$~erg s$^{-1}$, more than an order of
magnitude lower than the estimated fast outflow energy.

In modelling spectral structure in the higher resolution RGS data
(Figure 2R),  photoionised line emission had to be convolved with a
gaussian ($\sigma\sim$25 eV at 0.6 keV), corresponding to velocity
broadening of  29000 km s$^{-1}$ (FWHM),  consistent with  the high
velocity flow.

Applying the 2001 model to the 2004 data, a good fit was obtained with
element abundances and ionisation parameters fixed, the spectral change
being largely modelled by an increase in the steep, weakly absorbed
continuum component. In contrast, a change in the covering factor
(modelled by a lower column) of ionised absorber on the primary power
law was not able -- alone -- to fit both data sets.

\section{Broader application of the new spectral model}

The Black Hole Winds model (King and Pounds 2003) provides a simple
physical basis whereby massive, high velocity outflows can be expected
in AGN accreting at or above the Eddington limit. In addition to
PG1211+143, this might apply to both PDS456 (Reeves \etal\ 2003) and to
the bright Seyfert 1 IC4329A, recently shown (Markowitz \etal\ 2006) to
exhibit a strongly blue-shifted Fe K absorption line indicating a highly
ionised outflow at v$\sim$0.1c. While relatively rare in the local 
universe such X-ray spectra could be common for luminous, higher
redshift AGN.  

If the outflow velocity in PG1211+143 equates to the escape velocity at
the launch radius R$_{launch}$ (from an optically thick photosphere or
radiatively  extended inner disc), v$\sim$0.14c corresponds to
R$_{launch}$ $\sim$ $50R_{\rm s}$  (where $R_{\rm s} = 2GM/c^2$ is the
Schwarzschild radius), or $3\times 10^{14}$~cm for PG1211+143.  The EPIC
data show significant flux variability in the harder (2--10 keV) band on
timescales of 2--3 hours (fig 1 in Pounds  \etal\ 2003)), compatible with
the above scale size relating to the primary (disc/corona) X-ray
emission region.
 
Observations of PG1211+143 over several days are needed to test the new
spectral model. The variability timescale of the secondary power law
will constrain the scale of the region where we predict the fast outflow
undergoes internal shocks. Detecting hard X-ray emission from PG1211+143
above $\sim$20 keV would also support the need for a continuum component
with photon index less steep than that in the single power law/
absorption models,   while deeper RGS spectra are needed to resolve the
broad emission line profiles indicated in the model fits. 

In summary, it seems clear that further studies of high accretion rate
AGN, such as PG1211+143, offer great potential for understanding
outflows and their potential effects on the intergalactic medium and
host galaxy growth.

\bigskip

\discuss{Kazuo Makishima}{Comment: Steep power law you are finding is
extremely interesting and reminds me of similar phenomenon that occurs
in Galactic black hole binaries when they get in the very high state. In
that case this component appears between $10$--$50$~keV, connecting the
hard tail and the soft disc component.}

\discuss{Giorgio Matt}{Can you exclude that the $0.7$--$2$~keV spectrum is
due to ionized reflection rather than absorption?}

\discuss{Kenneth Pounds}{The spectral curvature (soft excess) in
PG1211 is too strong to be produced by photoionized reflection from the
accretion disc unless the direct continuum is somehow suppressed, for
example by light bending in strong gravity near the black hole. Given
the detection of a large column density outflow in this luminous AGN,
absorption plus re-emission provides a more likely explanation.}

\end{document}